\documentclass{article}
\usepackage{spconf,amsmath,graphicx, subcaption, float}
\usepackage[hidelinks]{hyperref}

\usepackage{xcolor}

\usepackage{enumitem}
\setlist{nosep, leftmargin=14pt}

\usepackage{mwe} 

\def\phfull{HArmonization BEnchmarking Tool}
\def\ph{HABET}

\def\numsiganova{n_F}
\def\fracsiganova{f_F}
\def\numsigttest{n_t}

\title{Harmonization Benchmarking Tool for Neuroimaging Datasets}

\name{
T. Osika$^{\star}$
\quad E. Ebrahim$^{\star}$
\quad M. Styner$^{\dagger}$
\quad M. Niethammer$^{\dagger}$
\quad T. Sawyer$^{\ddagger}$
\quad A. Enquobahrie$^{\star}$
}

\address{
$^{\star}$Kitware, Inc., NC, USA \\
$^{\dagger}$University of North Carolina at Chapel Hill, NC, USA \\
$^{\ddagger}$Elmhurst University, IL, USA
}

\begin{document}
%
\maketitle
\begin{abstract}

A major data pre-processing step for large, multi-site studies is to handle site effects by harmonizing data,
generating a dataset that enables more powerful analyses and more robust algorithms.
There is a wide variety of data harmonization techniques, but there are few tools that streamline the process of
harmonizing data,
comparing across techniques,
and benchmarking new techniques.
In this paper, we introduce \phfull{} (\ph{}), an open source tool for generating harmonized images and evaluating the performance of different harmonization algorithms. To demonstrate the capabilities of \ph{}, we harmonize diffusion MRI images from the Adolescent Brain and Cognitive Development (ABCD) study using two different approaches, and we compare their performance.

\end{abstract}
\section{Introduction} 

With the increasing collaboration and pooling of data among neuroimaging research centers,
the ability to reliably conduct mega-analyses is becoming more important.
Traditionally, meta-analyses have often been the only option, being possible even when only aggregated data is available from each study.
Mega-analysis,
in which unaggregated data is pooled across studies,
can be more powerful for analyzing structural neuroimaging data under typical conditions \cite{mega_better},
and it is increasingly an option with multi-site neuroimaging studies becoming more common \cite{biobank,adni,hagler_image_2019}.
A major challenge
when working with multi-site neuroimaging data
is to deal with effects attributable to study site,
such as
scanner hardware, reconstruction algorithms, acquisition protocols, and image quality.
These can all affect the final image \cite{pinto_harmonization_2020}
as well as downstream results \cite{vbm_site_effects_matter,dti_site_effects_matter}.
\emph{Data harmonization}
is the process of removing these effects.
When working to discover biomarkers or construct
predictive models out of multi-site medical image data,
harmonization is a 
critical 
preprocessing step to ensure that results can be generalized
and applied across a variety of clinical and research settings.


There is a wide variety of approaches to harmonization. A systematic review of harmonization literature was conducted in \cite{nan_data_2022}, where the authors classified existing techniques into four broad categories: distribution based techniques, image processing, synthesis, and invariant feature learning.
Distribution based techniques such as ComBat \cite{johnson_adjusting_2007} model the bias between different groups and subtract that bias out from the original data.
Image processing based techniques such as resampling, image filtering, and normalization, employ image processing algorithms to harmonize the data.
Synthesis and invariant feature learning methods typically involve deep learning.
Synthesis techniques aim to generate samples from missing groups, similar to the task of style transfer.
For example, this could transfer all of the images from different sites to match the image data distribution of a particular site.
Lastly, invariant feature learning methods involve encoding image data in some site-invariant form, and using that encoding instead of the original images in downstream tasks.

A closely related task to harmonization is \emph{domain adaptation}.
This term is sometimes used to refer to the invariant feature learning \cite{bento, adapt} or synthesis \cite{domainatm} categories described above,
but it typically refers to the broader problem of transforming data to match a given data distribution, beyond the context of multi-center studies.

The Adolescent Brain and Cognitive Development (ABCD) Study \cite{hagler_image_2019} is a large, multi-site, longitudinal study monitoring the brain and cognitive development of a national cohort of children.
Several studies have performed harmonization on the ABCD dataset.
In \cite{nielson_detecting_2018}, site effects in the ABCD fMRI images were examined by training a multinomial logistic classifier to predict scanner from the imaging data before and after ComBat harmonization \cite{johnson_adjusting_2007}.
In \cite{liu_learning_2021}, ABCD 2.0 T1 and T2 weighted images were harmonized with a deep learning approach. In \cite{cetin_karayumak_harmonization_2022}, ABCD DMRI images were harmonized with rotation invariant spherical harmonics (RISH).


Although many harmonization techniques have been developed,
there is a need for tools that streamline the process of
evaluating and comparing the various techniques on a given dataset.
In this work, we introduce the open source \textit{\phfull{}} (\ph),
and we apply it to investigate site effects in the
ABCD Study dataset.
This tool will be available to the research community as open source software;
code can be found at
\url{https://github.com/KitwareMedical/habet}.
Currently \ph{} supports ComBat and global scaling out of the box, and it
features a simple framework for adding new harmonization techniques. In principle,
\ph{} could support techniques from any of the four aforementioned categories except
for invariant feature learning, as these techniques skip the reconstruction step altogether.
More techniques will be made available in the future.
Report generation features allow researchers to quickly summarize and 
visualize the performance of different techniques.
Unlike domain adaptation libraries such as \cite{adapt}, our goal is
to provide \emph{evaluation metrics} for the effectiveness of site effect removal.
A similar work to ours is \cite{domainatm},
where the extent of harmonization is evaluated by comparing features
or images from a domain-adapted source dataset with those from a target dataset.
\ph{} instead focuses its evaluation on statistical model based detection of site effects with different types of harmonization, and it accommodates harmonization models that operate on more than two sites.
Our contributions are as follows:
\begin{enumerate}
    \item Introduce the open source framework \ph{} for harmonization benchmarking.
    \item Exemplify the capabilities of \ph{} by analyzing site effects in the ABCD data, before and after applying two different harmonization techniques.
\end{enumerate}

\section{Method}

\begin{figure}[htb]

  \centering
  \centerline{
    \includegraphics[scale=1]{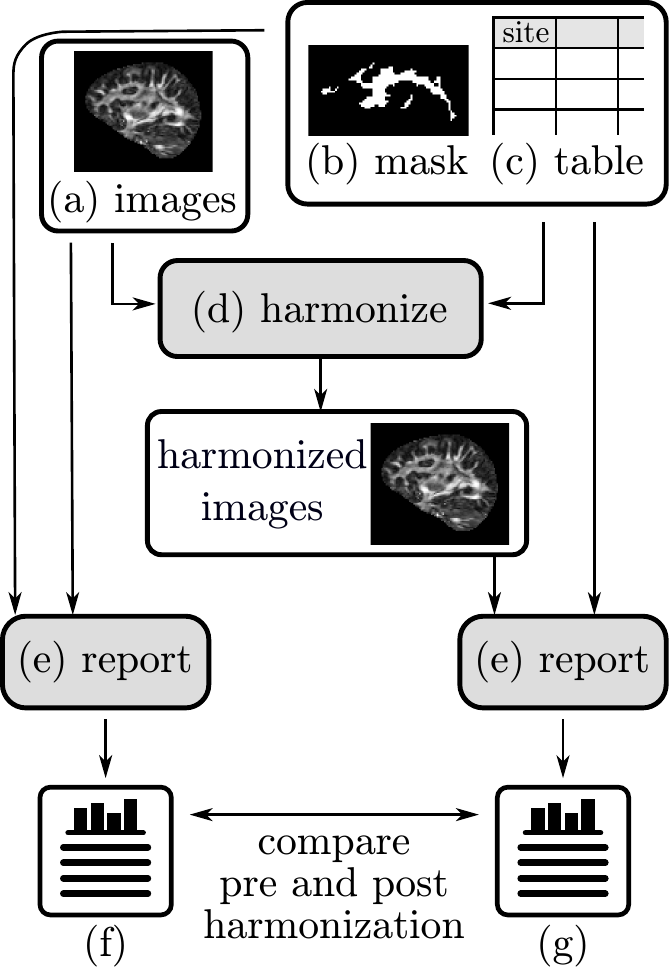}
  }

%
\caption{\label{fig:cfd}The
workflow to evaluate a harmonization technique. The input data are an image dataset (a), an optional mask (b) selecting voxels for inclusion, and a table (c) associating images to groups (such as study site) over which to remove effects. Our tool provides two modules: harmonization (d) and report generation (e). A harmonization technique is configured (d) and an analysis of site effects is conducted (e), producing reports of the significance and magnitude of site effects before (f) and after (g) harmonization.
}
\end{figure}

\ph{} provides a simple interface for both harmonizing imaging data and evaluating the performance of the harmonization technique used. \ph{} features two main components, harmonization and report generation; see an overview of the workflow in Figure~\ref{fig:cfd}. These two components are independent, so it is possible to use \ph{} only for the purpose of harmonizing images, or for comparing between different harmonization methods with already harmonized data.

Both components operate on a given set of images,
a mask selecting which part of the image to process,
and a table relating each image to some variables.
This includes the categorical variables whose effect on the image is undesirable
(e.g. site, scanner manufacturer), which we will refer to collectively as ``site.''
The table may also include covariates that are of interest for later analysis
and that could be confounded with site.
As previously stated, global scaling and ComBat are supported out of the box, with more harmonization methods to be added in the future.
\ph{} also features a simple API for registering custom harmonization techniques,
making it convenient for benchmarking newly developed approaches.

The report generation component aids in quantifying site effects for a given set of images. This can be executed on a set of images before and after harmonization, giving the user insight into the performance of a given set of harmonization methods.
At any given voxel location,
if there were no site effects then we would expect the within-site means
of the voxel values to be roughly equal, with some variation due to sampling
and other variables.
This hypothesis can be tested via one-way ANOVA:
we compute an $F$-statistic and an associated $p$-value at each voxel,
with site as the independent variable and voxel intensity as the dependent variable.
If, for a given voxel, the results of the ANOVA are found to be significant,
then we conclude that there are site effects at that image location.
We may wish to know, for those voxels that show significant $F$-scores,
\emph{which} sites cause the greatest discrepancy,
and for this we run a $t$-test for each pair of sites.
We report, as another measure of the extent of site effects,
the fraction of site pairs that showed significant effects
(similar to the approach in \cite{zav}).
Bonferroni correction is applied to the $F$-test $p$-values to account for the number of voxels,
and also to the $t$-test $p$-values to account for the number of voxels and site pairs.
Finally, we want a measure of site effect \emph{magnitude} when a location shows significant site effects. For this we report eta-squared values to indicate the proportion of the variance of each voxel value that can be attributed to site.
Putting it together, each report consists of:
\begin{enumerate}
    \item Site effect significance: a table of voxel-wise $F$-test results from the ANOVA and an image indicating the subset of the mask where site effects were detected.
    \item Site effect size: a table and an image of eta-squared values at each voxel.
    \item Pairwise site effect significance: a table of voxel-wise $t$-test results and an image depicting the fraction of significant $t$-tests at each location that showed significant site effects.
    \item Pairwise site effect size: a table of Hedges' $g$ values for voxels that showed significant site effects.
    \item A summary table reporting the the total numbers and fractions of $F$-tests and $t$-tests that were significant.
\end{enumerate}

\section{Results}


The ABCD study has 21 participating study sites across the country.
A wide range of data are collected, including structural and functional neuroimaging.
We focus on diffusion tensor imaging (DTI),
both as an illustration of our tool and as a precursor for our future development of tools for white matter microstructure analysis.
Ninety-six diffusion directions were used to acquire the DMRI images in the ABCD dataset, and the diffusion imaging protocols over the three different scanner manufacturers were chosen to be as similar as possible across sites \cite{hagler_image_2019}. For details on DMRI acquisition parameters and preprocessing, see \cite{hagler_image_2019}.
To examine site effects in the ABCD dataset, we took a sample of 836 subjects from the baseline year. We ensured that the values of site, baseline age, sex, scanner hardware, and scanner software version that occur in the sample were representative of the full ABCD dataset.
Deformable registration to align local structures was achieved with a deep learning approach
based on inverse consistency \cite{icon}.
All FA images were nonlinearly registered to the ``most typical'' subject image, in the sense of \cite{smith_tract-based_2006}:
the subject image selected for use as a template is the one with the minimum mean displacement when registered to all other images.
This approach lets us use a detailed and population-specific subject template, which improves the quality of the information available to the registration algorithm and can result in better alignment. To limit our analysis to white matter, a tissue type segmentation was computed from the T1-weighted MRI image of the subject image that was chosen as a common registration target.
This was done using FSL for brain extraction \cite{fsl_be} and tissue segmentation \cite{fsl_seg}.
The mask was then eroded to improve generalizability across the other images.

The images were harmonized using two approaches: ComBat and global scaling. Reports were generated for each technique and the original images using \ph{}.

\def\y{y}
\def\by{\bar{\y{}}}
\def\tl{\theta^\text{loc}}
\def\ts{\theta^\text{scl}}
\def\htl{\hat{\theta}^\text{loc}}
\def\hts{\hat{\theta}^\text{scl}}
\def\hbeta{\hat{\beta}}
\def\ep{\epsilon}

Global scaling fits a simple linear model to the site-averaged images, without allowing for spatial heterogeneity of site effects.
Let $\by{}_{iv}$ denote the value at voxel $v$ averaged over all the images in site $i$,
and let $\by{}_v$ denote the value at voxel $v$ averaged over \emph{all} the images.
The global scaling model, presented in \cite{fortin_harmonization_2017}, is
\begin{equation*}\label{eq:gs}
\by{}_{iv} = \tl_i + \ts_i\, \by_v + \ep_{iv},
\end{equation*}
where $\ep_{iv}\sim\mathcal{N}(0,\sigma^2)$.
After fitting the data to obtain the parameters $\tl_i$, $\ts_i$, and $\sigma^2$ at each voxel, harmonization proceeds by applying the transformation
$y\mapsto \frac{y-\htl_i}{\hts_i}$ to the values of each image from site $i$.
In the terminology of \cite{pinto_harmonization_2020}, this is a \emph{global} as opposed to \emph{voxel-wise} approach to harmonization. We use global scaling as a baseline global method to compare with voxel-wise methods like ComBat.

ComBat, introduced for genomics in \cite{johnson_adjusting_2007} and used for imaging data in \cite{fortin_harmonization_2017}, is based on the model
\begin{equation}\label{eq:combat}
\y{}_{ijv} = \alpha_v + X_{ij}\beta_v + \gamma_{iv} + \delta_{iv} \ep_{ijv},
\end{equation}
where $j$ indexes the subject images. Here the site effects are modeled by the parameters $\gamma_{iv}$ and $\delta_{iv}$;
see \cite{fortin_harmonization_2017} for details. 

When applying ComBat to the ABCD data, we included age and sex as covariates in the model (i.e. the matrix $X_{ij}$ in (\ref{eq:combat})), protecting these variables from having their effects washed out by the harmonization. We used a significance level of $\alpha = 0.05$ for the ANOVA and $t$-tests.

\begin{figure}[htb]

  \centering
  \centerline{
    \includegraphics[width=\linewidth]{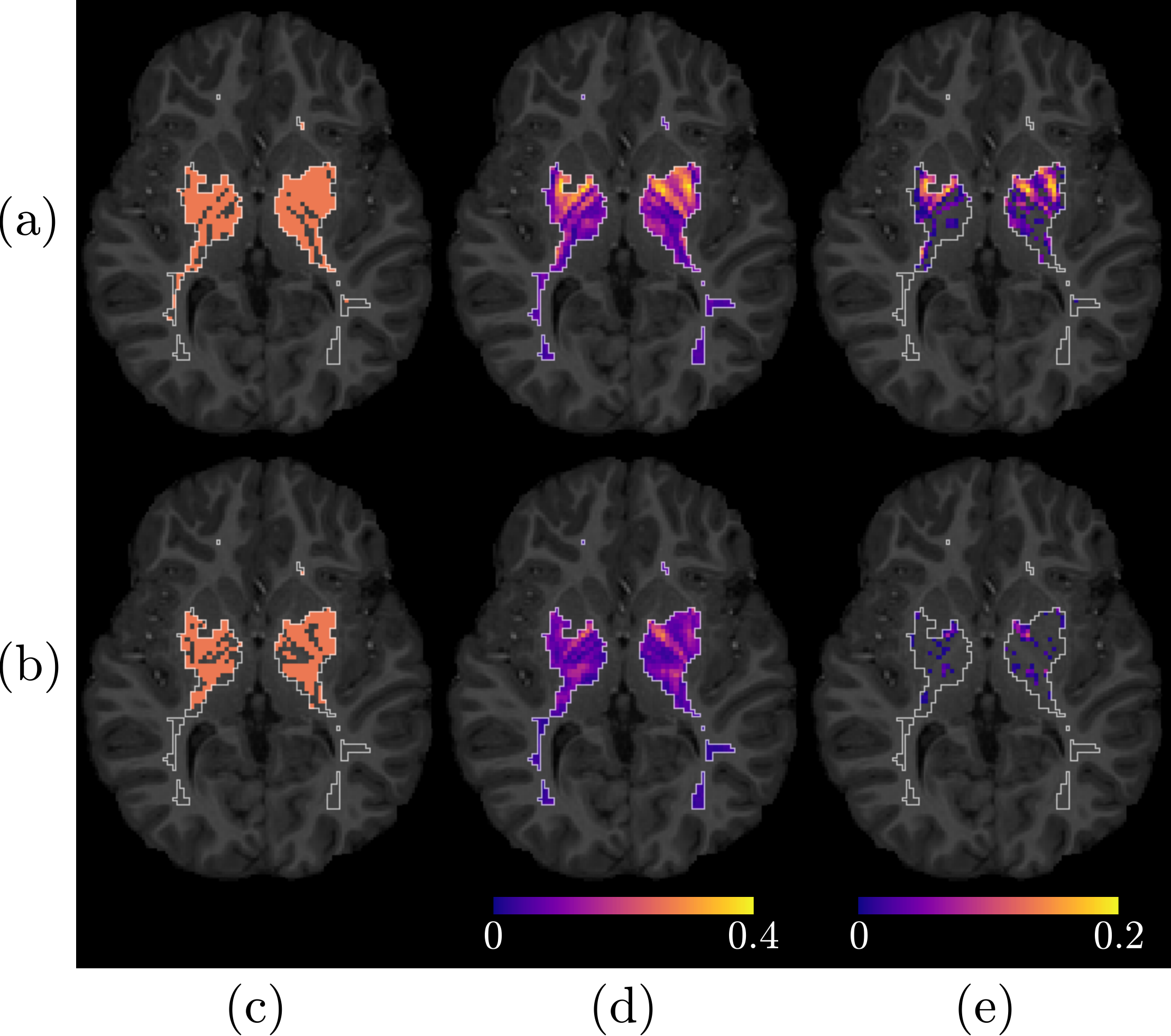}
  }

%
\caption{\label{fig:slices}Axial slices of the spatial
site effect report for
(a) unharmonized data and
(b) data harmonized wtih global scaling.
The values shown in color are
(c) binary ANOVA significance with Bonferroni corrected $\alpha=0.05$,
(d) eta-squared, and
(e) the fraction of $t$-tests that were significant with Bonferroni corrected $\alpha=0.05$.
No results are shown here for ComBat harmonized data,
because no voxels were flagged for significant site effects
after ComBat harmonization.
The white outline encloses the voxels that were selected for analysis,
and the background image is the T1-weighted MRI depicting the
common space to which all images were registered.
}
\end{figure}

\begin{figure}[htb]

  \centering
  \centerline{
    \includegraphics[width=\linewidth]{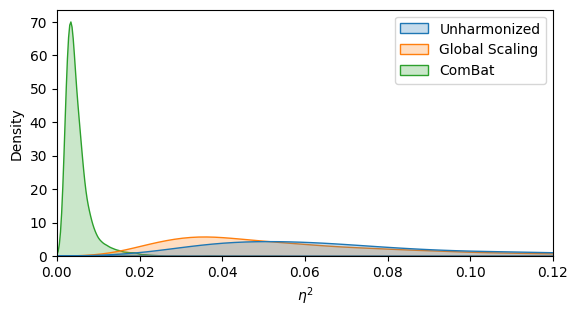}
  }

%
\caption{\label{fig:etasquared}The estimated distributions of $\eta ^ 2$ values
for the unharmonized images and the images harmonized with global scaling and ComBat.
Each density function was estimated from $\eta^2$ observations at each of the 11880 voxel locations under consideration. Note that all voxels are included, rather than only the
ones that had a significant ANOVA.
}
\end{figure}

Table~\ref{tbl:results} shows summary results for harmonized and unharmonized data: the number $\numsiganova{}$ and fraction $\fracsiganova{}$ of significant ANOVA tests and the number $\numsigttest{}$ of significant $t$-tests. Figure~\ref{fig:slices} shows a slice of each output image from the generated reports. For the unharmonized and global scaling images, all of the $\eta ^ 2$ values from the significant ANOVA tests were found to be above 0.06. Figure~\ref{fig:etasquared} shows the full distribution of $\eta ^ 2$ values, regardless of significance, for each harmonization method and for the unharmonized data.
\begin{center}
\begin{table}\centering
\begin{tabular}{| c |c | c | c |} 
 \hline
 Harmonization Method & $\numsiganova{}$ & $\fracsiganova{}$ & $\numsigttest{}$\\ [0.5ex] 
 \hline\hline
 None (Original Data) & 5123 & 0.43 & 24216 \\ 
 \hline
 Global Scaling & 3266 & 0.27 & 4887 \\
 \hline
 ComBat & 0 & 0 & N/A \\
 \hline
\end{tabular}
\caption{\label{tbl:results} The values of $\numsiganova{}$, $\fracsiganova{}$, and $\numsigttest{}$ reported for the ABCD sample before and after harmonization with global scaling and ComBat. Note that after harmonization with ComBat, we can no longer detect any site effects. N/A is reported for the value of $\numsigttest{}$ for ComBat because no $t$-tests were run.}
\end{table}
\end{center}

\section{Discussion}
We have demonstrated how \ph{} can be used to measure site effects in the ABCD diffusion MRI data and to compare the performance of different harmonization techniques.

In the unharmonized images, site effects were detected in nearly half (43\%) of the voxels chosen for analysis. All of the corresponding effect-sizes were above 0.06, which is typically considered to be a medium effect size. In short, site effects appear to be common and of considerable magnitude in the unharmonized images.

Global scaling decreased the percentage of voxels at which site effects were detected to 27\%. This is an improvement from the unharmonized images, but it is still high. Again, all of the corresponding effect-sizes were above 0.06. With ComBat, no site effects were detected. In \cite{fortin_harmonization_2017}, a similar vanishing of detected site effects was observed with ComBat when analyzing 156 ROI of the brain.

There are several areas in which our analysis could be improved.
First, Bonferroni correction is a very conservative method of multiple testing correction, partly because it assumes that
the statistical tests at various voxels are all independent.
Image values are typically highly correlated in space,
so the independence assumption is very far from reality.
This biases our analysis heavily towards false negatives
(i.e. failing to detect existing site effects).
When we report ``0 voxels with significant site effects'' after ComBat harmonization,
this could be a statement about the specificity of our site effect detector as much as it could be a statement about the effectiveness of ComBat.
We can get around this limitation by incorporating spatially informed techniques for multiple testing correction.

Second, our one-way ANOVA relies on the assumption that the intensity distribution at a given voxel is normally distributed within each site, with the same variance across sites.
The departure of the data from these assumptions is currently not reported.

The functionality of \ph{}  may be augmented by
introducing better methods for multiple testing correction,
supporting more image formats,
and supporting more harmonization methods out of the box.
Finally, we hope to make the tool more convenient
by providing easy access to a default dataset for benchmarking new techniques.

\section{Conclusion}
In this paper, we presented a new harmonization benchmarking tool, \ph{},
which can be used to harmonize a multi-site dataset, compare different harmonization techniques, and evaluate new harmonization techniques.
We demonstrated the use of \ph{} by
investigating site effects in DMRI images from the ABCD study dataset and
analyzing the effectiveness of ComBat compared to a simpler harmonization approach.


\section{Compliance with ethical standards}
\label{sec:ethics}

This research study was conducted retrospectively using data made available by the NIMH Data Archive. Ethical approval was not required because analyses using de-identified ABCD Study data are not considered human subjects research by NIH criteria.

\section{Acknowledgments}
\label{sec:acknowledgments}

This research was supported in part by the National Institutes of Health under Award Number R42MH118845. The content is solely the responsibility of the authors and does not necessarily represent the official views of the National Institutes of Health.

Data used in the preparation of this article were obtained from the ABCD Study (\url{https://abcdstudy.org}), held in the NIMH Data Archive (NDA). The data used for the analysis in this work and the raw results obtained can be found at DOI
\texttt{10.15154/1528197}.

\bibliographystyle{IEEEbib}
\bibliography{refs}

\end{document}